\documentclass{ifacconf}

\usepackage{amsmath,amsfonts,amssymb}
\usepackage{graphicx}      
\usepackage{graphics}
\usepackage{natbib}        

\newcommand{\be}{\begin{equation}}
\newcommand{\ee}{\end{equation}}
\newcommand{\bd}{\begin{displaymath}}
\newcommand{\ed}{\end{displaymath}}
\newcommand{\bea}{\begin{eqnarray}}
\newcommand{\eea}{\end{eqnarray}}
\newcommand{\Hi}{{\cal H}_\infty}
\newcommand{\Lxi}{\mathcal{L}_{\sigma}}
\newcommand{\LxiN}{\mathcal{L}_{\sigma}^N}
\newcommand{\R}{\mathbb{R}}
\newcommand{\C}{\mathbb{C}}

\newcommand{\Rp}{\mathbb{R}_0^{+}}
\newcommand{\psa}{pseudospectral abscissa{ }}
\newcommand{\ps}{pseudospectra }
\newcommand{\w}{\omega}
\newcommand{\alpf}{\alpha_f(\sigma)}
\newcommand{\Fs}{F_\sigma^{-1}}

\makeatletter
\def\Ddots{\mathinner{\mkern1mu\raise\p@
\vbox{\kern7\p@\hbox{.}}\mkern2mu
\raise4\p@\hbox{.}\mkern2mu\raise7\p@\hbox{.}\mkern1mu}}
\makeatother

\begin{document}
\begin{frontmatter}

\title{Computing the Pseudospectral Abscissa of Time-Delay Systems}

\author[First]{Suat Gumussoy},
\author[First]{Wim Michiels}

\address[First]{Department of Computer Science, K. U. Leuven, \\
        Celestijnenlaan 200A, 3001, Heverlee, Belgium \\
        \mbox{(e-mails: suat.gumussoy@cs.kuleuven.be, wim.michiels@cs.kuleuven.be)}.}

\begin{abstract}
The pseudospectra of a linear time-invariant system are the sets in the
complex plane consisting of all the roots of the characteristic equation
when the system matrices are subjected to all possible perturbations
with a given upper bound. The pseudospectral abscissa are defined as the
maximum real part of the characteristic roots in the pseudospectra and,
therefore, they are for instance important from a robust stability point
of view. In this paper we present a numerical method for the
computation of the pseudospectral abscissa of retarded delay
differential equations with discrete pointwise delays. Our approach is based on the connections between
the pseudospectra and the level sets of an appropriately defined complex
function. These connections lead us to a bisection algorithm for the
computation of the pseudospectral abscissa, where each step relies on
checking the presence of imaginary axis eigenvalues of an appropriately
defined operator. Because this operator is infinite-dimensional a
predictor-corrector approach is taken. In the predictor step the
bisection algorithm is applied where the operator is discretized into a
matrix, yielding approximations for the pseudospectral abscissa. The
effect of the discretization is fully characterized in the paper. In the
corrector step, the approximate pseudospectral abscissa are corrected to
any given accuracy, by solving a set of nonlinear equations that
characterize extreme points in the pseudospectra contours.
\end{abstract}


\end{frontmatter}

\section{Introduction}
The pseudospectra provide information about the characteristic roots of the system when the system matrices in the characteristic equation are subject to perturbations. They are closely related with the robust stability of the system and distance to instability, \cite{Trefethen:97}. We consider the characteristic equation of the time-delay systems:
\be \label{eq:chareqn}
\det F(\lambda)=0
\ee where
\be \label{eq:F}
F(\lambda):=\lambda I_{n\times n}-\left(\sum_{i=0}^m A_ie^{-\lambda\tau_i}\right),
\ee $A_i\in\C^{n\times n}$, $\tau_i\in\Rp$ for $i=0,\ldots,m$ and $\tau_0=0$. The maximum real part of the characteristic  roots is the \emph{spectral abscissa},
\be \label{eq:sa}
\alpha_0=\sup_{\lambda\in\C}\{\Re(\lambda): \det F(\lambda)=0 \}.
\ee
When the system matrices in (\ref{eq:F}) is subject to the perturbations, the \emph{pseudospectra} of the characteristic equation (\ref{eq:chareqn}) is defined as
\be \label{eq:psF1}
\Lambda_\epsilon=\left\{\lambda\in\C: \det\left(F(\lambda)+\Delta F(\lambda)\right)=0 \right\}
\ee where perturbations on the systems matrices are represented as
\be \label{eq:dF}
\Delta F(\lambda):=-\left(\sum_{i=0}^m \delta A_i e^{-\lambda \tau_i} \right),
\ee
$\delta A_i\in\C^{n \times n}$ and satisfying $\|\delta A_i\|_2\leq\frac{\epsilon}{w_i}$ for \mbox{$i=0,\ldots,m$}. Here the $w_i\in\Rp$  \mbox{$i=0,\ldots,m$} are some weights on the perturbations which can be chosen apriori.
The maximum real part in the \ps is the \emph{\psa} which is defined as
\be \label{eq:psa1}
\alpha_\epsilon=\sup_{\lambda\in\C}\{\Re(\lambda): \lambda\in\Lambda_\epsilon(F) \}.
\ee
The computation of the \psa for finite dimensional systems corresponds to the special case of (\ref{eq:chareqn}):
\bd
F_0(\lambda)=\lambda I_n-A_0.
\ed In this particular case, the \ps can be equivalently expressed as \cite{Boyd:89}
\be
\Lambda_\epsilon^0=\left\{\lambda\in\C: \sigma_{\max}\left(F_0^{-1}(\lambda)\right) >\frac{1}{\epsilon}\right\}
\ee where $\sigma_{\max}(A)$ is the largest singular value of the matrix $A$. Note that this definition reduces the \ps boundary to the level set of the resolvent norm. This connection is used to compute the distance to instability and \psa via a bisection algorithm in \cite{Byers:88} and \cite{Burke:03} respectively. A quadratically convergent algorithm for the \psa computation is given in \cite{BurkeCC:03} based on a `criss-cross' procedure. In \cite{Michiels:06}, these results are extended to matrix functions of the form (\ref{eq:F}) where the perturbations take the form of (\ref{eq:dF}). In particular, it is shown in Theorem $1$ that the \psa as defined in (\ref{eq:psF1}) can be expressed in the following way:
\be \label{eq:psF2}
\Lambda_\epsilon=\left\{\lambda\in\C: f(\lambda) >\frac{1}{\epsilon}\right\}
\ee where
\be
\label{eq:f} f(\lambda)=w(\lambda)\sigma_{\max}(F^{-1}(\lambda)) ,\ \ w(\lambda)=\sum_{i=0}^m \frac{|e^{-\lambda\tau_i}|}{w_i}.
\ee

Using the \ps definition in (\ref{eq:psF2}), the \psa in (\ref{eq:psa1}) can be rewritten as
\be \label{eq:psa2}
\alpha_\epsilon=\sup_{\lambda\in\C}\left\{\Re(\lambda): f(\lambda)=\frac{1}{\epsilon} \right\}.
\ee The supremum in the definition (\ref{eq:psa2}) is well-defined since $F^{-1}(\lambda)$ is a strictly proper function and $w(\lambda)$ is uniformly bounded on any right half complex plane in (\ref{eq:f}).

In Section~\ref{sec:psa}, a bisection algorithm is given for \psa computation of time-delay systems based on the connection between the \ps and the level sets of the function $f(\lambda)$.

This algorithm is implemented in two steps: first the approximate \psa is computed by the prediction step in Section~\ref{sec:psaapprox} and then the approximate results are corrected in Section~\ref{sec:psacorr}.

The overall algorithm for the \psa computation is outlined in Section \ref{sec:alg}. A numerical example and concluding remarks can be found in Sections~\ref{sec:ex} and \ref{sec:conc}.

\textbf{Notation:} \\
The notation in the paper is standard and given below.
\begin{tabular}{rl}
  $\C, \R:$ & the field of the complex and real numbers, \\
  $\Rp:$ & the positive real numbers including zero, \\
  $\Re(u):$ & real part of the complex number $u$, \\
  $\Im(u):$ & imaginary part of the complex number $u$. \\
  $|u|:$ & magnitude of the complex number $u$. \\
  $\bar{u}:$ & conjugate of the complex number $u$. \\
  $A^{*}:$ & complex conjugate transpose of the  \\
  & matrix $A$. \\
  $I_{n\times n}:$ & identity matrix with dimension $n$. \\
  $\sigma_{\max}(A):$ & the largest singular value of the matrix $A$. \\
  $\|F(j\w)\|_\infty:$ & $\mathcal{L}_\infty$ norm of the transfer function $F$
\end{tabular}

\section{The Bisection Algorithm for the Pseudospectral Abscissa Computation} \label{sec:psa}
Given the function $f$ in (\ref{eq:f}), define the function $\alpha_f(\sigma)$ as:
\be \label{eq:alpf}
\alpha_f(\sigma):=\sup_{\w\in\R}f(\sigma+j\w)
\ee over $\sigma\in(\alpha_0,\infty)$.

\begin{figure}
\begin{center}
\includegraphics[width=7cm]{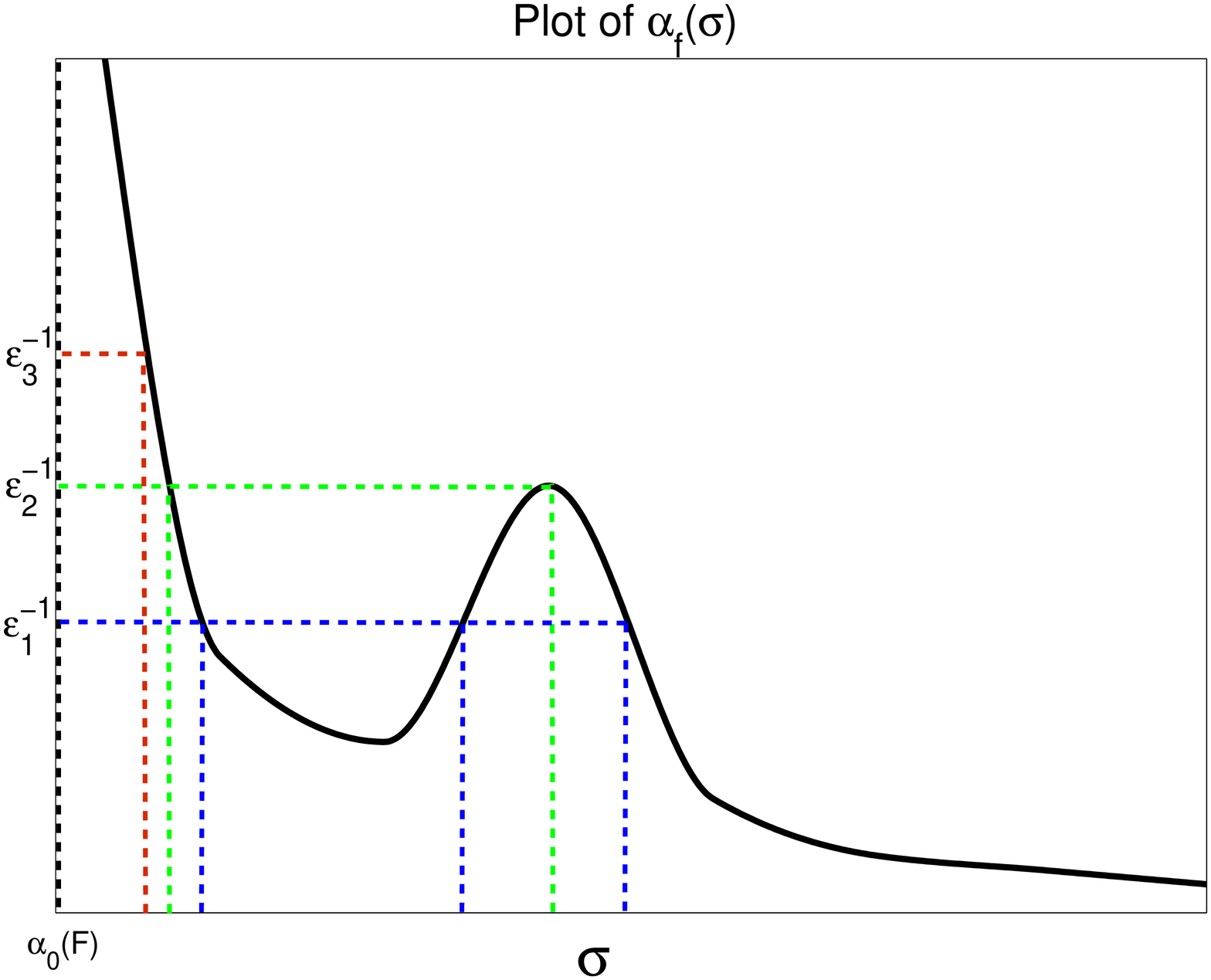}
\caption{\label{fig:prop1} $\alpf$ plot}
\end{center}
\end{figure}

\begin{prop} \label{prop:alpfmonot}
The function $\alpha_f(\sigma)$ is strictly monotonically decreasing over the interval $\sigma\in(\alpha_0,\infty)$.
\end{prop}
\noindent\textbf{Proof.\ } The proof is by contradiction. Assume that the function $\alpf$ is strictly increasing for some $\sigma$. Then there exists a level set $\frac{1}{\epsilon_1}$ such that there are at least two disjoint \ps regions in the complex plane due to (\ref{eq:psF2}) and the strictly properness of $F^{-1}(\lambda)$  (see the blue lines in Figure~\ref{fig:prop1}). Since $\alpf$ is a continuous function, one of the disjoint sets in the pseudospectra disappears without merging to other \ps sets for a higher level set $\frac{1}{\epsilon_3}$ (red lines in the Figure). This is a contradiction with the continuity of the individual eigenvalues and the fact that $F(\sigma)^{-1}$ is strictly proper (preventing eigenvalues to move off to infinity).

Similarly, it can be shown that the case $\alpha_f(\sigma_1)=\alpha_f(\sigma_2)$ with $\sigma_0<\sigma_1<\sigma_2$ is a contradiction with the continuity of the individual eigenvalues. \hfill $\Box$

\begin{prop} \label{prop:alpfinit}
The function $\alpha_f(\sigma)$ satisfies
\bea
\nonumber \lim_{\sigma\rightarrow\alpha_0(F)^{+}} \alpf &=& \infty, \\
\nonumber \lim_{\sigma\rightarrow +\infty} \alpf &=& 0.
\eea
\vspace{-0.5cm}
\end{prop}
\noindent\textbf{Proof.\ } The first assertion in the Proposition follows from the fact that there are characteristic roots on the boundary $\Re(s)=\alpha_0$. Therefore, the $\Hi$ norm of \mbox{$F^{-1}(\sigma+j\w)$} diverges to infinity since the denominator becomes singular. The second assertion is the result of that $F^{-1}$ is a strictly proper function. \hfill $\Box$

Using the definition in (\ref{eq:psa2}), the \psa is the $\sigma$ value of the function where $\alpf=\frac{1}{\epsilon}$. Since the function $\alpf$ is strictly decreasing by Proposition~\ref{prop:alpfmonot} and it attains all the values from $0$ to $\infty$ by Proposition~\ref{prop:alpfinit}, the \psa can be calculated by the following bisection algorithm.

\begin{alg} \label{alg:bisection} \ \
\begin{enumerate}
  \item[1)] $\sigma_L=\alpha_0$, $\sigma_R=\infty$, $\Delta \sigma=$tol,
  \item[2)] while $(\sigma_R-\sigma_L)>2\times\textrm{tol}$
  \begin{enumerate}
  \item[2.1)] $\Delta \sigma = 2 \times \Delta \sigma$,
  \item[2.2)] if $(\sigma_R=\infty)$ \\
              then $\sigma_M=\sigma_L+\Delta \sigma$, \\
              else $\sigma_M=\frac{\sigma_L+\sigma_R}{2}$.
  \item[2.3)] determine if $\alpha_f(\sigma_M)>\frac{1}{\epsilon}$ \\
              then $\sigma_L=\sigma_M$, \\
              else $\sigma_R=\sigma_M$.
  \end{enumerate}
   \item[] \hspace*{-0.8cm}\{result: the approximate \psa, \mbox{$\tilde{\sigma}=\sigma_L$}\}
\end{enumerate}
\end{alg}

The main computation in the bisection algorithm is to check whether the inequality in step $2.3$ is satisfied. By algebraic computation, the function $\alpf$ is equivalent to
\bea
\nonumber \alpf&=&\sup_{\w\in\R}f(\sigma+j\w), \\
\nonumber &=&\sup_{\w\in\R}\left\{w(\sigma+j\w)\sigma_{\max}(F^{-1}(\sigma+j\w))\right\}, \\
\nonumber &=&w(\sigma)\sup_{\w\in\R}\left\{\sigma_{\max}(F^{-1}(\sigma+j\w))\right\}, \\
&=&w(\sigma)\|F_\sigma^{-1}\|_\infty
\eea where
\be \label{eq:Fs}
F_\sigma^{-1}(j\w)=\left(j\w I_{n\times n}-\left(\sum_{i=0}^m A_{\sigma,i} e^{-j\w\tau_i}\right)\right)^{-1}
\ee and
\bea
\nonumber A_{\sigma,0}&=& A_0 - \sigma I_{n\times n}, \\
A_{\sigma,i}&=& A_i e^{-\tau_i \sigma},\ \textrm{for}\ i=1,\ldots,m.
\eea

The inequality $\alpha_f(\sigma)>\frac{1}{\epsilon}$ in step $2.3$ is satisfied if and only if  $F_\sigma^{-1}(j\w)$ has a singular value equal to $\frac{1}{\epsilon w(\sigma)}$ for some value of $\w$. This condition can be reduced into the verification of the imaginary axis eigenvalues of an infinite dimensional operator as shown in the following Theorem.

\begin{thm} \label{thm:GLxi}
The inequality
\be \label{eq:ineq:Lxi}
\alpha_f(\sigma)>\frac{1}{\epsilon}
\ee is satisfied if and only if the linear infinite dimensional operator $\Lxi$ has eigenvalues on the imaginary axis where $\Lxi$ is defined on $X:=\mathcal{C}([-\tau_{\max},\ \tau_{\max}],\mathbb{C}^{2n})$ by
\begin{eqnarray}
 \mathcal{D}(\mathcal{L}_{\sigma}) =\left\{\phi\in X:\
\phi^{\prime}\in X,\hspace{3.5cm} \right. \label{def:Lxi2}\\
\nonumber \hspace{0.8cm} \phi^{\prime}(0)=M_{0}\phi(0)  + \sum_{i=1}^m
(M_i\phi(-\tau_i)+M_{-i}\phi(\tau_i) ) \},
\end{eqnarray}

\begin{equation}
\mathcal{L}_{\sigma}\phi=\phi^{\prime},\;\phi\in\mathcal{D}(\Lxi) \label{def:Lxi}
\end{equation}
with
\[
\begin{array}{l}
M_{0}=\left[\begin{array}{cc}  A_{\sigma,0} & -(\epsilon w(\sigma))^2 I_{n\times n} \\
-I_{n\times n} & -A_{\sigma,0}^*
\end{array}\right],\\
M_i=\left[\begin{array}{cc} A_{\sigma,i} &0\\0&0
\end{array}\right],\ \
M_{-i}=\left[\begin{array}{cc} 0 &0\\0&-A_{\sigma,i}^*
\end{array}\right],\ \ 1\leq i\leq N.
\end{array}
\]
\end{thm}
\noindent\textbf{Proof.\ } A similar proof is given in \cite{Gumussoy:09}. This Theorem generalizes Proposition $28$ of \cite{Genin:02}.  \hfill $\Box$

Using Theorem~\ref{thm:GLxi}, we can refine the \emph{conceptual} algorithm for the \psa computation,

\begin{alg} \label{alg:bisectLxi} \ \
\begin{enumerate}
  \item[1)] $\sigma_L=\alpha_0$, $\sigma_R=\infty$, $\Delta \sigma=$tol,
  \item[2)] while $(\sigma_R-\sigma_L)>2\times\textrm{tol}$
  \begin{enumerate}
  \item[2.1)] $\Delta \sigma = 2 \times \Delta \sigma$,
  \item[2.2)] if $(\sigma_R=\infty)$ \\
              then $\sigma_M=\sigma_L+\Delta \sigma$, \\
              else $\sigma_M=\frac{\sigma_L+\sigma_R}{2}$.
  \item[2.3)] if $\mathcal{L}_{\sigma_M}$ has imaginary axis eigenvalues \\
              then $\sigma_L=\sigma_M$, \\
              else $\sigma_R=\sigma_M$.
  \end{enumerate}
   \item[] \hspace*{-0.8cm}\{result: the approximate \psa, \mbox{$\tilde{\sigma}=\sigma_L$}\}
\end{enumerate}
\end{alg}

Note that step $2.3$ in the Algorithm \ref{thm:GLxi} requires solving a linear infinite dimensional eigenvalue problem for $\Lxi$ which needs to be discretized in a practical implementation. We will do this using a \emph{spectral method} (see, e.g. \cite{Trefethen:00,Breda:05,Breda:06}) and calculate the approximate solution by solving the standard linear eigenvalue problem. This approach is described in the next section.

\section{Predicting the \psa} \label{sec:psaapprox}

Given a positive integer $N$, we consider a mesh $\Omega_N$ of $2N+1$ distinct points in the interval $[-\tau_{\max},\ \tau_{\max}]$:
\begin{equation}\label{defmesh}
\Omega_N=\left\{\theta_{N,i},\ i=-N,\ldots,N\right\},
\end{equation}
where
\[
\hspace{-2.2cm} -\tau_{\max}\leq\theta_{N,-N}<\ldots<
\theta_{N,-1}<\theta_{N,0}=0
\\
\]
\[
\hspace{4.5cm} <\theta_{N,1}<\cdots<\theta_{N,N}\leq\tau_{\max}.
\]
This allows to replace the continuous space $X$ with the space $X_N$
of discrete functions defined over the mesh $\Omega_N$, i.e.\ any
function $\phi\in X$ is discretized into a block vector
$x=[x_{-N}^T\cdots\ x_{N}^T ]^T\in X_N$ with components
\[
x_i=\phi(\theta_{N,i})\in\mathbb{C}^{2n},\ \ i=-N,\ldots,N.
\]
Let $\mathcal{P}_N x,\ x\in X_N$ be the unique $\mathbb{C}^{2n}$
valued interpolating polynomial of degree $\leq 2N$ satisfying
\[
\mathcal{P}_N x (\theta_{N,i})=x_{i},\ \ i=-N,\ldots,N.
\]
In this way, the operator $\mathcal{L}_{\sigma}$ over $X$ can be
approximated with the matrix $\mathcal{L}_{\sigma}^N:\ X_N\rightarrow
X_N$, defined as

\begin{equation}\label{defldisc}
\begin{array}{ll}
\left(\mathcal{L}_{\sigma}^N\ x\right)_i=&\left(\mathcal{P}_N
x\right)^{\prime}(\theta_{N,i}),\quad i=-N,\ldots,-1,\\
\left(\mathcal{L}_{\sigma}^N\ x\right)_0=
&M_0 \mathcal{P}_N x(0)+\sum_{i=1}^m (M_i \mathcal{P}_N x(-\tau_i)+ M_{-i} \mathcal{P}_N x(\tau_i))
\\
\left(\mathcal{L}_{\sigma}^N\ x\right)_i=&\left(\mathcal{P}_N
x\right)^{\prime}(\theta_{N,i}), \quad i=1,\ldots,N.
\end{array}
\end{equation}

Using the Lagrange representation of $\mathcal{P}_N x$,
\[
\begin{array}{l}
\mathcal{P}_N x=\sum_{k=-N}^N l_{N,k}\ x_k,\ \ \
\end{array},
\]
where the Lagrange polynomials $l_{N,k}$ are real valued polynomials
of degree $2N$ satisfying
\[
l_{N,k}(\theta_{N,i})=\left\{\begin{array}{ll}1 & i=k,\\
0 & i\neq k,
\end{array}\right.
\]
 we obtain the explicit form
\[
\mathcal{L}_{\sigma}^N=
\left[\begin{array}{lll}
d_{-N,-N} &\hdots & d_{-N,N} \\
\vdots & & \vdots \\
d_{-1,-N} &\hdots & d_{-1,N} \\
a_{-N} & \hdots & a_N\\
d_{1,-N} &\hdots & d_{1,N} \\
\vdots & & \vdots \\
d_{N,-N} &\hdots & d_{N,N}
\end{array}\right]\in\mathbb{R}^{(2N+1)(2n)\times(2N+1)2n},
\]
where
\[
\begin{array}{lll}
d_{i,k}&=&l^{\prime}_{N,k}(\theta_{N,i}) I,\ \ \ \
i,k\in\{-N,\ldots,N\},\;i\neq0\\
a_0&=& M_0\ x_0+\sum_{i=1}^m \left(M_i
l_{N,0}(-\tau_i)+M_{-i}l_{N,0}(\tau_i)\right) \\
a_{k}&=&\sum_{i=1}^m \left(M_i
l_{N,k}(-\tau_i)+M_{-i}l_{N,k}(\tau_i)\right) \\
&& \hspace{3.5cm} k\in\{-N,\ldots,N\},\ k\neq0.
\end{array}
\]

Note that all the problem specific information and the parameter $\sigma$ are concentrated in the middle row of $\mathcal{L}_{\sigma}^N$, i.e.\ the elements $(a_{-N},\ldots,a_N)$, while all other elements of $\mathcal{L}_{\sigma}^N$ can be computed beforehand.

Since step $2.3$ of Algorithm \ref{alg:bisectLxi} is based on checking the presence
of eigenvalues of $\Lxi$ on the imaginary axis and thus strongly rely on the symmetry of the eigenvalues with respect to the imaginary axis, it is important that this property is \emph{preserved} in the discretization. The following Proposition gives the condition on the mesh such that this symmetry holds.

\begin{prop}\label{propsymLxi}
If the mesh $\Omega_{N}$ satisfies
\begin{equation}\label{meshSymmetry}
\theta_{N,-i}=-\theta_{N,i},\ i=1,\ldots,N,
\end{equation}
then the following result hold:
for all $\lambda\in\mathbb{C}$, we have
\begin{equation}\label{symextraLxi}
\det\left(\lambda I - \LxiN\right)=0 \Leftrightarrow \det\left(-\bar{\lambda}-\LxiN\right)=0.
\end{equation}
\end{prop}

\noindent\textbf{Proof.\ } Consider the differentiation matrix
with elements
\[
\Delta_{k,l}=l'_{N,k-N-1}(\theta_{N,l-N-1}),\ \
k,l\in\{1,\ldots,2N+1\}.
\]
and let $U$ be such that $U^{-1}\Delta U=\Delta^T$.
Define the matrix $S\in\R^{2N+1\times2N+1}$ with terms equal to $1$ on the main skew diagonal and $0$ elsewhere, the symmetry property of the mesh (\ref{meshSymmetry}) assures that
\[
\left(  U^{-1}S\right)\otimes
\left[\begin{array}{cc} 0&I_n\\I_n&0\end{array}\right]
 \mathcal{L}^N_{\sigma} \left(U S\right)\otimes
\left[\begin{array}{cc}
0&I_n\\I_n&0\end{array}\right]=-\left(\mathcal{L}_{\sigma}^N\right)^*,
\]
that is, the matrices $\LxiN$ and
$-\left(\LxiN\right)^T$ are similar.
 The proposition directly follows. \hfill $\Box$

Based on the discretization of $\Lxi$ into $\LxiN$, we propose the following algorithm to approximate (predict) the pseudospectral abscissa. It corresponds to Algorithm \ref{alg:bisectLxi} where step $2.3$ is replaced with the matrix $\LxiN$ for a fixed $N$.

\begin{alg} \label{alg:bisectLxiN} \ \
\begin{enumerate}
  \item[1)] $\sigma_L=\alpha_0$, $\sigma_R=\infty$, $\Delta \sigma=$tol,
  \item[2)] while $(\sigma_R-\sigma_L)>2\times\textrm{tol}$
  \begin{enumerate}
  \item[2.1)] $\Delta \sigma = 2 \times \Delta \sigma$,
  \item[2.2)] if $(\sigma_R=\infty)$ \\
              then $\sigma_M=\sigma_L+\Delta \sigma$, \\
              else $\sigma_M=\frac{\sigma_L+\sigma_R}{2}$.
  \item[2.3)] if $\mathcal{L}_{\sigma_M}^N$ has imaginary axis eigenvalues \\
              then $\sigma_L=\sigma_M$, \\
              else $\sigma_R=\sigma_M$.
  \end{enumerate}
   \item[] \hspace*{-0.8cm}\{result: the approximate \psa, \mbox{$\tilde{\sigma}=\sigma_L$}\}
\end{enumerate}
\end{alg}

In what follows we clarify the effect of using the discretized operator in the algorithm. The next Theorem establishes the link between the imaginary axis eigenvalues of $\LxiN$ and the corresponding inequality check condition similar to the connection between $\Lxi$ and the inequality (\ref{eq:ineq:Lxi}) in Theorem~\ref{thm:GLxi}.
\begin{thm} \label{thm:Lxi-GN}
Assume that the mesh $\Omega_N$ is symmetric around the zero as given in (\ref{meshSymmetry}). Let $p_N$ be the polynomial of the degree $2N$ satisfying the conditions,
\begin{eqnarray}
p_N(0;\ \lambda)&=&1, \\
p_N^{\prime}(\theta_i;\lambda)&=&\lambda p_N(\theta_i;\lambda),\;
\nonumber i=-N,\ldots,-1,1,\ldots,N.
\end{eqnarray}
The matrix $\LxiN$ has an imaginary axis eigenvalue $\lambda=j\omega$ if and only if the inequality
\be \label{eq:ineqLxiN}
\alpha_f^N(\sigma)>\frac{1}{\epsilon}
\ee holds where
\begin{eqnarray}\label{defGN}
\nonumber \alpha_f^N(\sigma)&:=&\sup_{\w\in\R}f_N(\sigma+j\w)\ \textrm{and} \\
\nonumber f_N(\sigma+j\w)&=&w(\sigma)\left(j\omega I-A_{\sigma,0}-\sum_{i=1}^m A_{\sigma,i} p_N(-\tau_i;\
j\omega)\right)^{-1}.
\end{eqnarray}
\end{thm}

Therefore, the effect of using $\LxiN$ instead of $\Lxi$ corresponds to computing the \emph{approximate} \psa
\bd
\alpha_\epsilon^N=\sup_{\lambda\in\C}\left\{\Re(\lambda): f_N(\lambda)=\frac{1}{\epsilon} \right\}.
\ed

The accuracy of the approximation depends on the discretization parameter $N$. Therefore, the accuracy can be chosen arbitrarily close to $\alpha_\epsilon$ by increasing $N$. Note that at each iteration of step $2.3$ in Algorithm~\ref{alg:bisectLxiN}, an eigenvalue problem of size $(2n)(2N+1)$ needs to be solved which may be computationally very demanding for large $N$.

However, because the eigenvalues of $\LxiN$ exhibit the spectral convergence to the corresponding eigenvalues of $\Lxi$ (following the lines of \cite{Breda:05}) and because the approximation error can be corrected in a very cheap way as we shall see in the next section, it is sufficient to have a small value of $N$ for most practical applications.

\section{Correcting the \psa} \label{sec:psacorr}

The Bisection Algorithm \ref{alg:bisectLxiN} finds the complex points $\tilde{\lambda}_i=\tilde{\sigma}+j\tilde{\w}_i$ for $i=1,\ldots,\tilde{n}$ achieving the approximate \psa $\tilde{\sigma}$ which is close to the $\alpha_\epsilon$ given tolerance and discretization points $N$. These approximate results are corrected by using the property that the eigenvalues of the \psa appear as solutions of a finite dimensional nonlinear eigenvalue problem. The following theorem establishes the link between this nonlinear eigenvalue problem and the linear eigenvalue problem of $\Lxi$.

\begin{thm} \label{thm:Lxi-Hxi}
$\lambda$ is an eigenvalue of linear operator $\Lxi$ if and only if
\begin{equation} \label{prob:HxiEigThm}
\det H_{\sigma}(\lambda)=0,
\end{equation}
where
\begin{eqnarray} \label{eq:HamMatrix}
H_{\sigma}(\lambda):=\lambda I-M_0&-&\sum_{i=1}^m \left(M_i
e^{-\lambda\tau_i}+M_{-i}e^{\lambda\tau_i}\right)
\end{eqnarray}
and the matrices $M_0$, $M_i$, $M_{-i}$ are defined in Theorem \ref{thm:GLxi}.
\end{thm}

The solutions of (\ref{prob:HxiEigThm}) can be
found by solving
\begin{equation}
H_{\sigma}(\lambda)\ v=0,\ \ \lambda\in\mathbb{C},\
v\in\mathbb{C}^{2n},\ v\neq 0, \label{prob:HxiEig}
\end{equation}
which in general has an infinite number of solution.

The correction method is based on the property that if \mbox{$\w(\sigma)\|\Fs(j\omega)\|_\infty=\frac{1}{\epsilon}$}, then the operator $\Lxi$, or equivalently, (\ref{prob:HxiEig}) has a multiple non-semisimple eigenvalue as shown in Figure~\ref{fig:semisimple}:

\begin{figure}
\begin{center}
\resizebox{8cm}{!}{\includegraphics{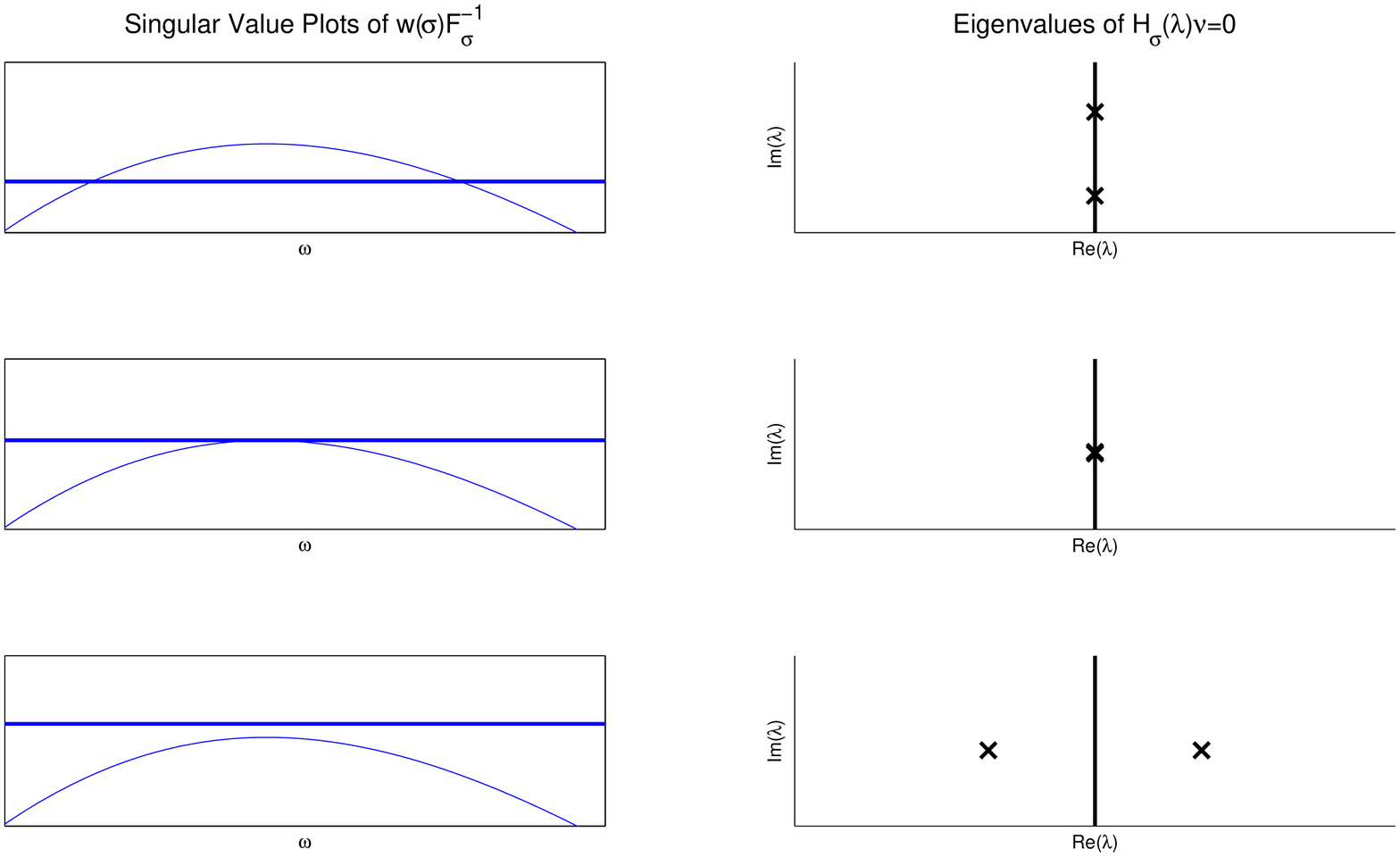}}
 \caption{\label{fig:semisimple} (left) Intersections of the
 singular value plot of $w(\sigma)F_\sigma^{-1}$ with the horizontal line $\frac{1}{\epsilon}$
 for the cases where (top) $w(\sigma)\|F_\sigma^{-1}(j\w)\|_\infty>\frac{1}{\epsilon}$, (middle) $w(\sigma)\|F_\sigma^{-1}(j\w)\|_\infty=\frac{1}{\epsilon}$ and
 (bottom) $w(\sigma)\|F_\sigma^{-1}(j\w)\|_\infty<\frac{1}{\epsilon}$. (right) Corresponding eigenvalues of the problem
 (\ref{prob:HxiEig}).}
\end{center}
\end{figure}

If $\lambda_\epsilon=\alpha_\epsilon+j\w_\epsilon$ are such that
\begin{equation}\label{direct0}
w(\sigma)\|\Fs(j\omega)\|_{\infty}=\frac{1}{\epsilon}=w(\alpha_\epsilon)\sigma_{\max}\left(F_{\alpha_\epsilon}^{-1}(j\w_\epsilon)\right),
\end{equation}
then setting
\[
h_{\sigma}(\lambda)=\det H_{\sigma}(\lambda),
\]
the pair $(\w,\alpha)=(\w_\epsilon,\alpha_\epsilon)$ satisfies
\begin{equation}\label{direct1}
h_{\sigma}(j\omega)=0,\ \ h_{\sigma}^{\prime}(j\omega)=0.
\end{equation}
These complex-valued equations seem over-determined but this is not
the case due to the spectral properties of $H_{\sigma}(\lambda)$. Using the symmetry of the eigenvalues of the nonlinear eigenvalue problem (\ref{prob:HxiEig}) with respect to imaginary axis, we can write the following:
\begin{cor}\label{corextra}
For $\omega\geq 0$, we have
\begin{equation}\label{col1}
\Im\ h_{\sigma}(j\omega)=0
\end{equation}
 and
\begin{equation}\label{col2}
 \Re\ h_{\sigma}^{\prime}(j\omega)=0.
\end{equation}
\end{cor}
\noindent\textbf{Proof.\ } From the symmetry property of the eigenvalues with respect to the imaginary axis,
\[
h_{\sigma}(\lambda)=h_{\sigma}(-\lambda),\ \ \ h_{\sigma}^{\prime}(
\lambda)=-h_{\sigma}^{\prime}( -\lambda).
\]
Substituting $\lambda=j\omega$ yields
\[
\begin{array}{l}
h_{\sigma}(j\omega)=h_{\sigma}(-j\omega)=\left(h_{\sigma}(j\omega)\right)^*, \\
h_{\sigma}^{\prime}(j\omega)=-h_{\sigma}^{\prime}(-j\omega)=-\left(h_{\sigma}^{\prime}(j\omega)\right)^*,
\end{array}
\]
and the assertions follow. \hfill $\Box$

\medskip

\noindent Using Corollary~\ref{corextra} we can simplify the
conditions (\ref{direct1}) to:
\begin{equation}\label{direct2x}
\left\{\begin{array}{l}
\Re\ h_{\sigma}(j\omega)=0 \\
\Im\ h_{\sigma}^{\prime}(j\omega)=0
\end{array}\right..
\end{equation}
Hence, the pair $(\w_\epsilon,\alpha_\epsilon)$  satisfying (\ref{direct0})
can be directly computed from the two equations (\ref{direct2x}),
e.g.\ using Newton's method, provided that good starting values are
available.

\medskip

\noindent The drawback of working directly with (\ref{direct2x}) is
that an explicit expression for the determinant of $H_{\sigma}$ is
required. To avoid this, let $u,v\in\mathbb{C}^n$ be such that
\be \label{eq:Hsig}
H_{\sigma}(j\omega) \left[\begin{array}{c}u\\
v\end{array}\right]=0,\ \ \ n(u,v)=0,
\ee
where $n(u,v)=0$ is a normalizing condition. Given the structure of
$H_{\sigma}$ it can be verified that a corresponding left eigenvector
is given by $[-v^*\ u^*]$. According to \cite{Lancaster:99}, we get
\[
h_{\sigma}'(j\omega)=0\Leftrightarrow [-v^*\ u^*]\ H^{\prime}_{\sigma}(j\omega) \left[\begin{array}{c}u\\
v\end{array}\right]=0.
\]
A simple computation yields:
\begin{equation}
[-v^*\ u^*]\ H_{\sigma}^{\prime}(j\omega) \left[\begin{array}{c}u\\
v\end{array}\right]=2\Im\left\{v^*\left(I+\sum_{i=1}^p A_{\sigma,i}\tau_i
e^{-j\omega\tau_i}\right)u\right\},
\end{equation}
which is always real. This is a consequence of the property
(\ref{col2}).

\medskip

\noindent Taking into account the above results, we end up with
$4n+3$ real equations
\begin{equation}\label{forfinal}
\left\{\begin{array}{l}
H_\sigma(j\omega,\ \sigma)\left[\begin{array}{c}u, \\
v\end{array}\right]=0, \quad n(u,v)=0\\
\Im\left\{v^*\left(I+\sum_{i=1}^p A_{\sigma,i}\tau_i e^{-j\omega\tau_i}\right)u\right\}=0\\
\end{array}\right.
\end{equation}
in  the $4n+2$ unknowns $\Re(v),\Im (v),\Re(u),\Im(u),\omega$ and
$\sigma$. These equations are still overdetermined because  the
property (\ref{col1}) is not explicitly exploited in the
formulation, unlike the property (\ref{col2}). However, it makes the
equations (\ref{forfinal}) exactly solvable, and the $(\omega,\sigma)$
components have a one-to-one-correspondence with the solutions of
(\ref{direct2x}).

In conclusion, as a result of the bisection algorithm in the prediction step, the approximate \psa $\tilde{\sigma}$ and the corresponding critical frequencies $\tilde{\w}_i$ for $i=1,\ldots,\tilde{n}$ are calculated. Note that these computations are based on the approximation of $\Lxi$ into a matrix $\LxiN$. Using these approximate results as estimates of $(\w_\epsilon,\alpha_\epsilon)$ (\ref{direct0}), we can compute the approximate eigenvectors $u$ and $v$. These approximate values improved in the correction step by solving (\ref{forfinal}). At the end of the correction step, the \psa $\sigma=\alpha_\epsilon$ and the achieved frequency $\omega=\omega_\epsilon$ are obtained within predefined tolerance.

\section{Algorithm} \label{sec:alg}

The overall algorithm for computing the \psa consists of two steps: the prediction step and the correction step. The first step requires a repeated computation of the eigenvalues of a $(2N+1)2n\times (2N+1)2n$ matrix $\LxiN$. The second step solves (\ref{forfinal}) with $4n+3$ equations and $4n+2$ unknowns using Gauss-Newton algorithm. Our method chooses $N$ sufficiently large such that the results of the prediction step are good starting values for the correction step. Note that by increasing $N$ and using only the prediction step, the approximate \psa can be computed arbitrarily close to $\alpha_\epsilon$. However, this approach has more numerical cost than the combined approach when $N$ is large.

\begin{alg} \label{alg:overall} \ \
${}$\\
\textit{Input:} system data, tolerance tol for prediction step, discretization points $N$\\
\textit{Output:} \psa $\alpha_\epsilon$ \\

\underline{\textit{Prediction Step}:}
\begin{enumerate}
  \item[1)] Calculate the spectral abscissa $\alpha_0$ of $F$ (\ref{eq:F}),
  \item[2)] $\sigma_L=\alpha_0$, $\sigma_R=\infty$, $\Delta \sigma =$tol,
  \item[3)] while $(\sigma_R-\sigma_L)>2\times\textrm{tol}$
  \begin{enumerate}
  \item[3.1)] $\Delta \sigma = 2 \times \Delta \sigma$,
  \item[3.1)] if $(\sigma_R=\infty)$ \\
              then $\sigma_M=\sigma_L+\Delta \sigma$, \\
              else $\sigma_M=\frac{\sigma_L+\sigma_R}{2}$.
  \item[3.2)] if $\mathcal{L}_{\sigma_M}^N$ has imaginary axis eigenvalues \\
              then $\sigma_L=\sigma_M$, \\
              else $\sigma_R=\sigma_M$.
  \end{enumerate}
   \item[] \hspace*{-0.8cm}\{result: the approximate \psa, \mbox{$\tilde{\sigma}=\sigma_L$} and the corresponding frequencies $j\tilde{\w}_i$ $i=1,\ldots,\tilde{n}$ of $\LxiN$\}
\end{enumerate}

\underline{\textit{Correction Step}:}
\begin{enumerate}
\item calculate the approximate null vectors $\left\{x_1,\ldots,x_{\tilde{n}}\right\}$ of $H_{\tilde{\sigma}}(j\tilde{\w}_i)$ $i=1,\ldots,\tilde{n}$,
 \item for all $i\in\{1,\ldots,\tilde{n}\}$, solve (\ref{forfinal}) with
 starting values
 \[
\left[\begin{array}{c}u\\ v\end{array}\right]=x_i,\
\omega=\tilde{\omega}_i,\ \ \sigma=\tilde{\sigma}
 \]
denote the solution with $(u_{\epsilon,i}, v_{\epsilon,i},\omega_{\epsilon,i},\sigma_{\epsilon,i})$.
\item set
$\alpha_\epsilon:=\max_{1\leq i\leq \tilde{n}} \sigma_{\epsilon,i}$.
\end{enumerate}
\end{alg}

In our implementation, mesh points are chosen as Chebyshev extremal points since the corresponding interpolation polynomial has less oscillation towards the end of the interval compared to another distribution of mesh points, \cite{Breda:06}.

Note that the spectral abscissa calculation in the prediction step requires the calculation of the right-most eigenvalue of time-delay systems. This computation is done by DDE-BIFTOOL, \cite{Engelborghs:02}. The overall algorithm is fully automated and implemented as a MATLAB function.

\section{Example} \label{sec:ex}

\begin{figure}
\begin{center}
\includegraphics[width=8cm]{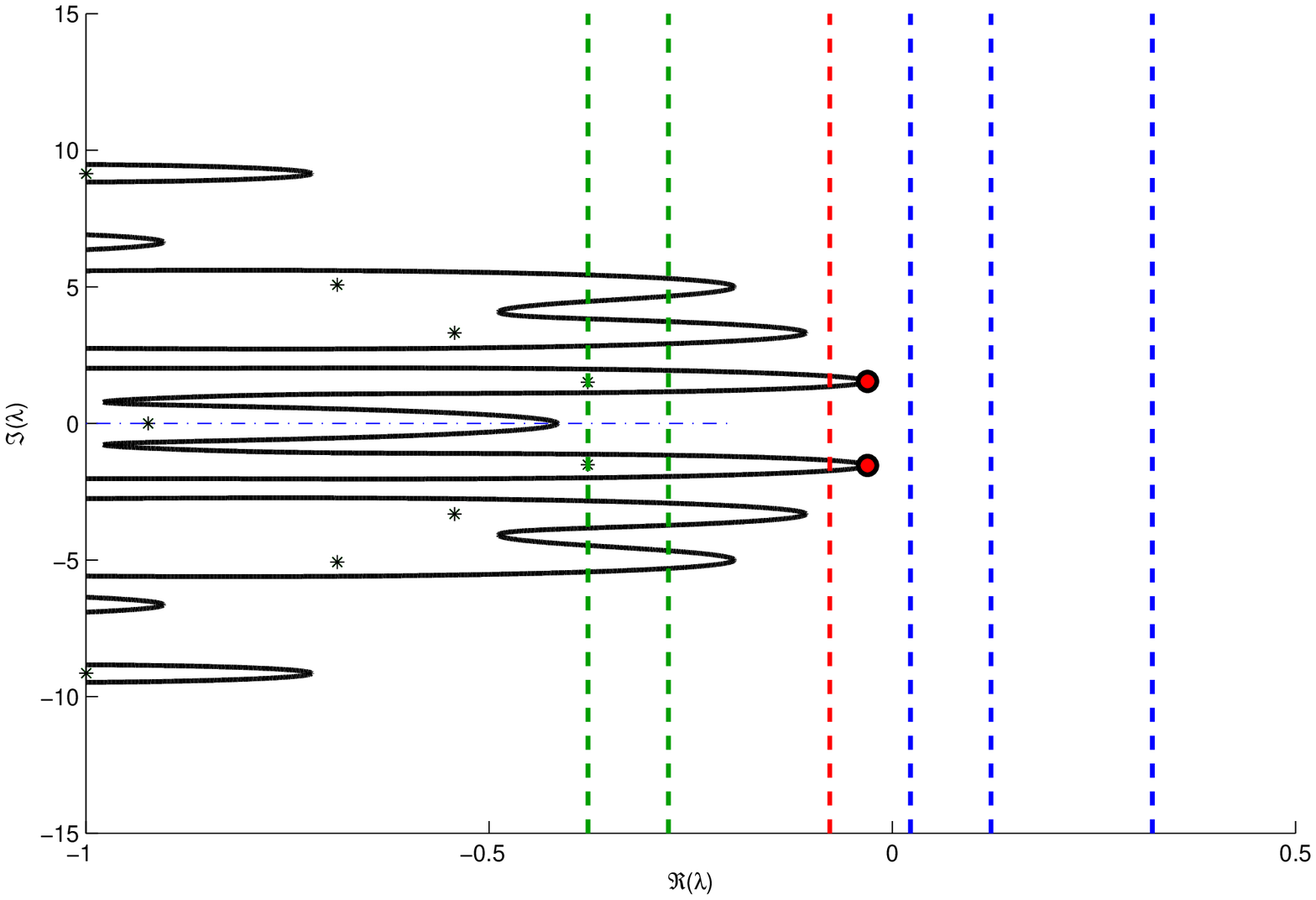}
\caption{\label{fig:example} The \ps and the \psa}
\end{center}
\end{figure}

We tested the numerical method on several benchmark problems. We generated the following difficult example to benchmark our method. We consider a time-delay system $F$ in (\ref{eq:F}) with the dimensions $m=7$, $n=10$, $n_u=2$, $n_y=4$ with delays $\tau_1=0.1$, $\tau_2=0.2$, $\tau_3=0.3$, $\tau_4=0.4$, $\tau_5=0.5$, $\tau_6=0.6$, $\tau_7=0.8$. The weights $w_i$ are set to $1$ and $\epsilon=0.1$. The \ps is shown with black lines and black stars indicate the characteristic roots of (\ref{eq:chareqn}) in Figure~\ref{fig:example}.

The tolerance in the bisection algorithm is set to $0.05$ and the discretization parameter is chosen as $N=6$. Each iteration of the while loop in the prediction step computes $\sigma_M$ and updates $\sigma_L$ or $\sigma_R$ shown as the vertical green and blue lines respectively. The approximate \psa as a result of the prediction step is $\tilde{\sigma}=-0.0774$ and the corresponding critical frequencies are $\tilde{\w}_1=1.3493$, $\tilde{\w}_2=1.7318$.

These approximate values are improved in the correction step and the computed \psa is $\alpha_\epsilon=-0.0307$ at $\w_\epsilon=1.5383$ shown as red dots in Figure~\ref{fig:example}.

\section{Concluding Remarks} \label{sec:conc}

An accurate method to compute the \psa of retarded time-delay systems with arbitrary number of delays is
given. The method is based on two steps: The prediction step calculates the approximate \psa using the connection between \ps and the level set of a function. The correction step computes the \psa by solving equations based on the nonlinear eigenvalue problem. The method is successfully applied to the moderate size example and its effectiveness is shown.

\begin{ack}
This article present results of the Belgian Programme on
Interuniversity Poles of Attraction, initiated by the Belgian State,
Prime Minister's Office for Science, Technology and Culture, and of
OPTEC, the Optimization in Engineering Centre of the K.U.Leuven.
\end{ack}

\bibliography{ifacconf}             
\end{document}